%% file: main.tex
\documentclass[sigconf,9pt]{acmart}

\copyrightyear{2023}
\acmYear{2023}
\setcopyright{acmlicensed}\acmConference[ICS '23]{2023 International Conference on Supercomputing}{June 21--23, 2023}{Orlando, FL, USA}
\acmBooktitle{2023 International Conference on Supercomputing (ICS '23), June 21--23, 2023, Orlando, FL, USA}
\acmPrice{15.00}
\acmDOI{10.1145/3577193.3593716}
\acmISBN{979-8-4007-0056-9/23/06}

\usepackage{algorithmic}
\usepackage{graphicx}
\usepackage{textcomp}

\usepackage{xcolor}
\def\BibTeX{{\rm B\kern-.05em{\sc i\kern-.025em b}\kern-.08em
    T\kern-.1667em\lower.7ex\hbox{E}\kern-.125emX}}

\input{header}
\begin{document}

\title{Revisiting Temporal Blocking Stencil Optimizations}

\newcommand{\tsc}[1]{\textsuperscript{#1}}


\author{Lingqi Zhang\tsc{1,3}, Mohamed Wahib\tsc{2}$^*$, Peng Chen\tsc{3,2}$^*$, Jintao Meng\tsc{4}, Xiao Wang\tsc{5}$^*$, Toshio Endo\tsc{1}}\thanks{$^*$ Corresponding authors}
\author{Satoshi Matsuoka\tsc{2,1}}
\affiliation{
  \institution{\tsc{1} Tokyo Institute of Technology, Japan, {\small\{zhang.l.ai@m.titech.ac.jp, endo@is.titech.ac.jp\}}}
  \institution{\tsc{2} RIKEN Center for Computational Science, Japan,  {\small\{mohamed.attia@riken.jp, matsu@acm.org\}}}
  \institution{\tsc{3} National Institute of Advanced Industrial Science and Technology, Japan, {\small\{chin.hou@aist.go.jp\}}}
  \institution{\tsc{4} Shenzhen Institutes of Advanced Technology, China, {\small\{jt.meng@siat.ac.cn\}}}
  \institution{\tsc{5} Oak Ridge National Laboratory, USA, {\small\{wangx2@ornl.gov\}}}
  \country{}
}

\renewcommand{\shortauthors}{Lingqi Z. et al.}
\renewcommand{\authors}{Lingqi Zhang, Mohamed Wahib, Peng Chen, Jintao Meng, Xiao Wang, Toshio Endo and Satoshi Matsuoka}

\begin{abstract}

Iterative stencils are used widely across the spectrum of High Performance Computing (HPC) applications. Many efforts have been put into optimizing stencil GPU kernels, given the prevalence of GPU-accelerated supercomputers. To improve the data locality, temporal blocking is an optimization that combines a batch of time steps to process them together. Under the observation that GPUs are evolving to resemble CPUs in some aspects, we revisit temporal blocking optimizations for GPUs. We explore how temporal blocking schemes can be adapted to the new features in the recent Nvidia GPUs, including large scratchpad memory, hardware prefetching, and device-wide synchronization. We propose a novel temporal blocking method, EBISU, which champions low device occupancy to drive aggressive deep temporal blocking on large tiles that are executed tile-by-tile. We compare EBISU with state-of-the-art temporal blocking libraries: STENCILGEN and AN5D. We also compare with state-of-the-art stencil auto-tuning tools that are equipped with temporal blocking optimizations: ARTEMIS and DRSTENCIL. Over a wide range of stencil benchmarks, EBISU achieves speedups up to $2.53$x and a geometric mean speedup of $1.49$x over the best state-of-the-art performance in each stencil benchmark. 
\end{abstract}

\maketitle


\input{01_introduction}
\input{02_background}

\input{03_EBISU}

\input{04_Implementation}

\input{05_AttainablePerformance}

\input{06_ImplementationDecisions}
\input{07_evaluation}
\input{08_relatedwork}
\input{09_conclusion}
\input{10_ack}
\bibliographystyle{ACM-Reference-Format}
\bibliography{acmart}
\end{document}

%% file: header.tex
\usepackage{mdframed}
\usepackage{lipsum}

\usepackage{algorithmic}
\usepackage{graphicx}
\usepackage{textcomp}
\usepackage{xcolor}
\usepackage{url}
\usepackage{hyperref}
\usepackage{enumitem}
\usepackage{soul}
\usepackage[outline]{contour}
\usepackage{hhline}
\usepackage{soul}

\setlist[itemize]{leftmargin=3.0mm}
\def\BibTeX{{\rm B\kern-.05em{\sc i\kern-.025em b}\kern-.08em
    T\kern-.1667em\lower.7ex\hbox{E}\kern-.125emX}}

\usepackage[skip=0pt]{caption}
\usepackage{colortbl}
\usepackage{threeparttable}
\input{myliststyle}
\usepackage{comment}
\usepackage{adjustbox}

\usepackage[switch]{lineno} 

\newcommand{\revision}[1]{{\color{black}{#1}}}
\newcolumntype{|}{!{\vrule width 0.5pt}}
\newcolumntype{?}{!{\vrule width 1pt}}
\newcolumntype{^}{!{\vrule width 1.2pt}}

\makeatother
\usepackage{multirow}
\usepackage{tablefootnote}
\usepackage{todonotes}
\usepackage{subcaption}

\definecolor{Gray}{gray}{0.9}

%% file: myliststyle.tex
\usepackage{listings}
\usepackage{color}
\definecolor{codegreen}{rgb}{0,0.6,0}
\definecolor{codegray}{rgb}{0.5,0.5,0.5}
\definecolor{codepurple}{rgb}{0.58,0,0.82}
\definecolor{backcolour}{rgb}{0.95,0.95,0.92}
 
\usepackage{caption}

\lstdefinestyle{mystyle}{
    backgroundcolor=\color{white},   
    commentstyle=\color{codegreen},
    keywordstyle=\color{magenta},
    numberstyle=\tiny\color{codegray},
    stringstyle=\color{codepurple},
    basicstyle=\footnotesize,
    breakatwhitespace=false,         
    breaklines=true,                 
    captionpos=b,                    
    keepspaces=true,                 
    numbers=left,                    
    numbersep=5pt,                  
    showspaces=false,                
    showstringspaces=false,
    showtabs=false, 
    tabsize=2
}
 

%% file: 01_introduction.tex
\section{Introduction}
Stencils are patterns in which a mesh of cells is updated based on the values of the neighboring cells. They are common computational patterns that exist widely in many scientific applications. They account for up to 49\% of workloads in many HPC centers~\cite{hagedorn2018high}. Applications of stencils include mainly finite difference solvers of Partial Differential Equations. (PDEs)~\cite{berger1984adaptive,datta2009optimization}. \revision{PDEs further support a wide spectrum of applications, spanning from weather modeling and seismic simulations to fluid dynamics simulations~\cite{10.1145/2749246.2749255}.}. 

Many efforts have gone into optimizing stencils~\cite{akbudak2020asynchronous,wellein2009efficient,Holewinski:2012:HCG:2304576.2304619}. Due to the low computational intensity of stencils~\cite{stengel2015quantifying}, combining steps and processing them together, i.e., temporal blocking, is an optimization widely used in iterative stencils~\cite{maruyama2011physis,malas2015multicore,endo2018applying,wellein2009efficient,zohouri2018combined}. This optimization increases the computational intensity, which comes at the price of adopting complex schemes to handle the constraints of temporal dependencies. Traditionally, temporal blocking resolves the dependency between time steps either by redundant overlapping of tiles~\cite{DBLP:conf/cgo/MatsumuraZWEM20,rawat2018domain,Holewinski:2012:HCG:2304576.2304619,Meng:2009:PMA:1542275.1542313} or by complicated titling geometry (e.g. diamond~\cite{bondhugula2017diamond} and hexagonal~\cite{grosser2014hybrid,grosser2014relation}). Either way, the overhead of resources for resolving the temporal blocking dependencies increases the data with the depth of temporal blocking~\cite{DBLP:conf/cgo/MatsumuraZWEM20}. An increasing number of time steps to block gradually moves the kernel's bottleneck from the memory throughput to be bound by either the memory latency or register pressure~\cite{DBLP:conf/cgo/MatsumuraZWEM20}. Among temporal blocking optimization efforts, many of them are related to specific hardware, e.g., FPGA~\cite{zohouri2018combined,waidyasooriya2016opencl}, CGRA~\cite{podobas2020template}, multi-core~\cite{datta2008stencil}, and many-core~\cite{rawat2018domain} architectures. We focus in this paper on GPUs due to their prevalence in HPC systems~\cite{top500}. 

When closely observing the latest GPUs~\footnote{We focus on Nvidia GPUs in this paper since the continuity of GPU products by Nvidia over decades provides the grounds for observing changes.}, there are notable changes in key features. We observe a significant increase in cache capacity. Specifically, the total capacity of the user-managed cache (shared memory) increased from $720$ KB in K20~\cite{kepler} to $17,712$ KB in A100~\cite{nvidiaa100}. The shared memory capacity has increased $24.6$x in recent decades. In addition, GPUs provide features that have been supported by CPUs for years. Examples include cooperative groups (i.e., device-wide barriers), low(er) latency of operations, and asynchronous copy of shared memory (i.e., prefetching)~\cite{guide2022cuda}. 

These new developments open opportunities for aggressive optimizations in stencil kernels. However, existing state-of-the-art temporal blocking implementations, e.g. AN5D~\cite{DBLP:conf/cgo/MatsumuraZWEM20} and STENCILGEN~\cite{rawat2018domain}, are designed to run at high occupancy and are hence relatively conservative in the use of resources to avoid adverse pressure on resources (ex: register spilling). For example AN5D~\cite{DBLP:conf/cgo/MatsumuraZWEM20} uses at maximum $96$ registers per thread and STENCILGEN~\cite{rawat2018domain} uses at maximum $64$ registers per thread for all the benchmarks reported. Yet the limit for registers is $255$ in both V100 and A100~\cite{guide2022cuda} GPUs. For shared memory usage, AN5D~\cite{DBLP:conf/cgo/MatsumuraZWEM20} consumes at most $34.8$ MB per thread block and STENCILGEN~\cite{rawat2018domain} uses at most $33.8$ MB per thread blocks. Yet the limit for shared memory is $164$ MB in A100~\cite{guide2022cuda} GPUs. This conservative manner is in part due to the intention for ensuring a higher occupancy. 

In this paper, we take inspiration from the work of Volkov et al.~\cite{volkov2010better}; we propose a different approach to occupancy and performance in temporal blocking. We first determine a parallelism setting that is minimal in occupancy while sufficient in instruction level parallelism. We base our approach for temporal blocking on lower occupancy, i.e., we build large tiles running at minimum possible concurrency to be executed tile-by-tile, and accordingly scale up the use of on-chip resources to run the tile at maximum possible performance.

We propose \emph{EBISU}: Epoch (temporal) Blocking for Iterative Stencils, with Ultracompact parallelism. EBISU's design principle is to run the code at the minimum possible parallelism that would saturate the device, and then use the freed resources to scale up the data reuse and reduce the dependencies between tiles. Though the idea is seemingly simple, the challenge is the lack of design principles to achieve scalable optimizations for temporal blocking. In other words, temporal blocking schemes in literature are designed to avoid pressure on resources since resources are scarce in over-subscribed execution; EBISU on the other hand assumes ample of resources that are freed due to running in low occupancy and the goal is to scale the data reuse to all the available resources for a single tile at a time that spans the entire device. We drive EBISU through a cost model that makes the decision on how to scale the use of resources effectively at low occupancy. 

The contributions of this paper are as follows:
\begin{itemize}
    \item We propose the design principle of EBISU: low-occupancy execution of a single-tile at a time while scaling the use of resources to improve \revision{data locality}.
    \item We include \revision{an analysis of the practical attainable performance } to support the design decisions \revision{for} EBISU. \revision{We build on our analysis to identify how various factors contribute to the performance of EBISU. }
    \item We evaluate EBISU across a wide range of stencil benchmarks. Our implementation achieves significant speedup over state-of-the-art libraries and implementation. We achieve a geomean speedup of 1.53x over the top performing state-of-the-art implementations for each stencil benchmark.
\end{itemize}

  

%% file: 02_background.tex
\begin{figure}[t]
\centering
 \includegraphics[width=\linewidth]{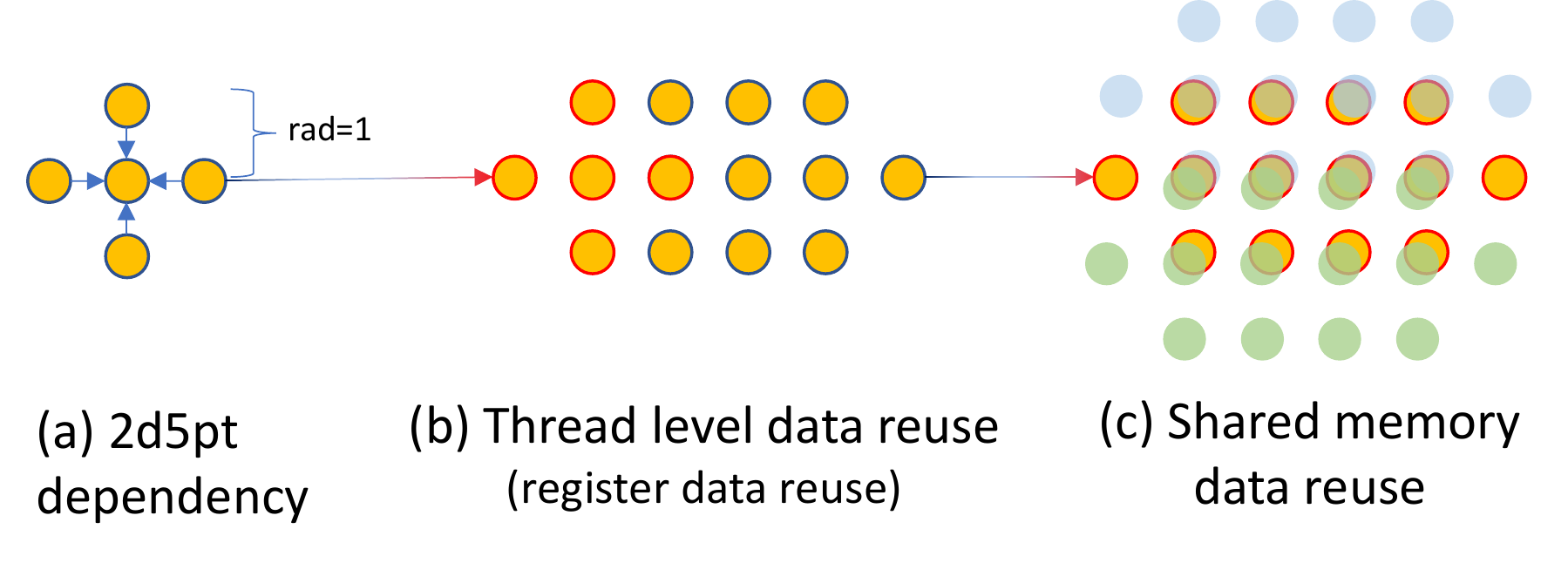}
 \caption{Spacial Blocking, using 2D 5-point Jacobian (2d5pt) stencil as an example}
 \label{fig:spblk}

\end{figure}
\begin{figure}[t]
\centering
 \includegraphics[width=\linewidth]{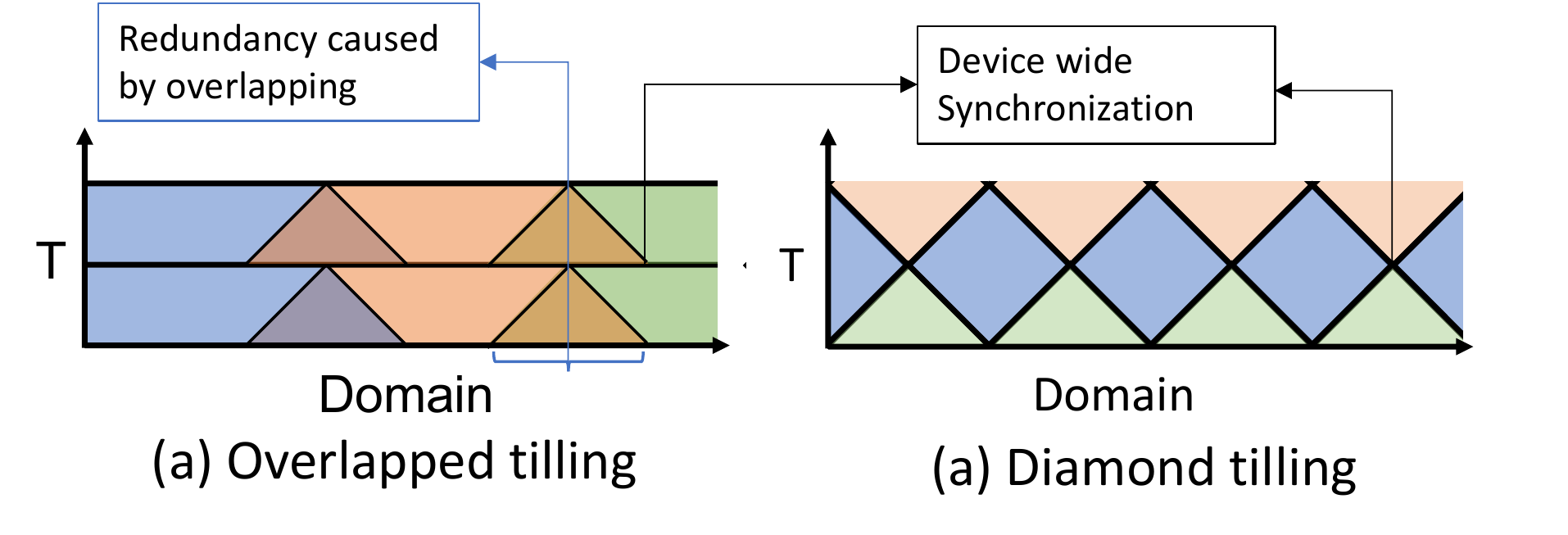}
 \caption{Temporal Blocking}
 \label{fig:tblk}
\end{figure}
\input{LISTING/samplestencils}


\begin{figure*}[ht]
\centering
\centering
\includegraphics[width=\textwidth]{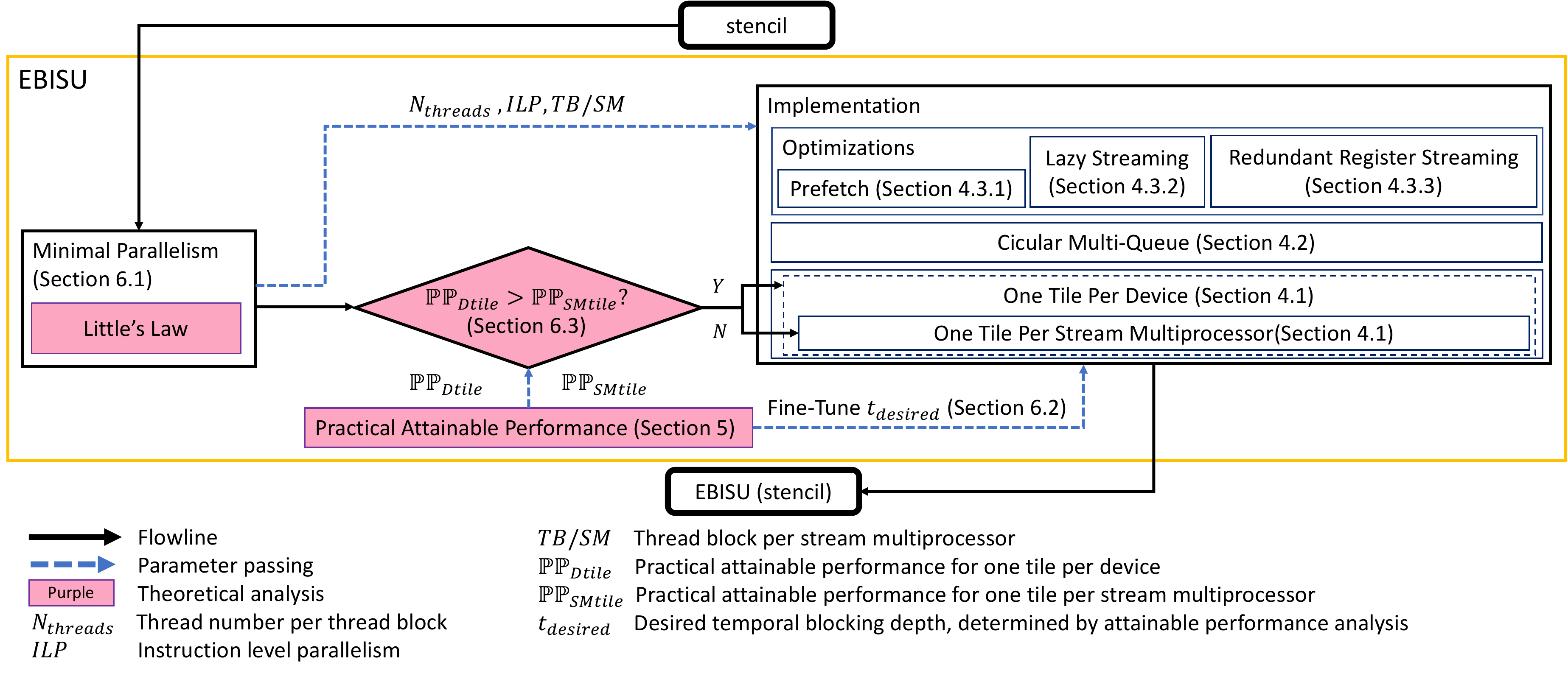}
\caption{\label{fig:overview}\revision Overview of EBISU. 
}
\end{figure*}
\section{Background}

\subsection{Stencils}\label{sec:stencilback}
Stencils are characterized by their memory access patterns. We present the pseudo code for the 1D 3-Point, 2D 5-Point and 3D-7-Point Jacobian Stencil in Listing~\ref{Fig:1d3p}, Listing~\ref{Fig:2d5p}, and Listing~\ref{Fig:3d7p} respectively. We use a 2D Jacobian 5-point (2d5pt) stencil as an example. Figure~\ref{fig:spblk}.a illustrates the neighborhood dependencies of the 2d5pt stencil. In order to compute one point, the four adjacent points are necessary. Two blocking methods are widely used to optimize iterative stencils for data locality:  

\subsubsection{Spatial Blocking}
In spatial blocking on GPUs, thread (blocks) load a single tile of the domain to its local memory to improve the data locality among adjacent locations~\cite{irigoin1988supernode,Wolfe:1989:MIS:76263.76337}. The local memory can be registers~\cite{zhao2019exploiting,chen2019versatile}(Figure~\ref{fig:spblk}.b) and scratchpad memory~\cite{maruyama2014optimizing,rawat2019optimizing}(Figure~\ref{fig:spblk}.c). However, halo layer(s) are still unavoidable.

\subsubsection{Temporal Blocking}
In iterative stencils, each time step depends on the result of the previous time step. An alternative optimization is to \revision{to combine several time steps to expose temporal locality}
~\cite{DBLP:conf/cgo/MatsumuraZWEM20,rawat2016effective}. In this case, the temporal dependency is resolved by overlapped tilling~\cite{Holewinski:2012:HCG:2304576.2304619,Meng:2009:PMA:1542275.1542313,10.1145/2830018.2830025} (Figure~\ref{fig:tblk}.a) or by applying complex geometry~\cite{10.1145/2458523.2458526,MURANUSHI20151303} (Figure~\ref{fig:tblk}.b, diamond tiling~\cite{grosser2014relation,bondhugula2017diamond} as an example). The main short-coming of overlapped tiling is redundant computation, while the main disadvantage of complex geometry is an adverse effects on cache hits~\cite{malas2015multicore}. Additionally, complex geometry is penalized by the device-wide synchronization necessary to ensure that the result is updated in the global memory. 

\subsubsection{N.5-D Temporal Blocking}
N.5-D blocking~\cite{DBLP:conf/cgo/MatsumuraZWEM20,rawat2018domain,nguyen20103} is a combination of spatial blocking and overlapped temporal blocking~\cite{nguyen20103}. Take 3.5-D temporal blocking as an example. We do spatial tiling in the X and Y dimensions, and then stream in the Z dimension (2.5-D spatial blocking). As we stream over the Z dimension, each XY plane would conduct a series of temporal steps (1-D temporal blocking). This method reduces the overhead of redundant computations in an overlapped temporal blocking schema. 

\subsection{GPU Architecture}
\subsubsection{CUDA Programming Model}
The CUDA programming model includes: the base execution unit, thread; 32 threads grouped into a block of schedule units, warp; Several warps grouped together into a thread block unit; and thread block units can be grouped into a grid. When mapping the programming model to the GPU architecture, the CUDA driver maps the thread block to a Stream Multiprocessor (SM) and grids to a GPU device. The mapping abides by the rules of dividing the resources among the threads. For example, at most 8 thread blocks and at most 2048 threads can reside concurrently on a stream multiprocessor. Also, the total number of registers and shared memory in a stream multiprocessor also limits the number of thread block that can run concurrently. 

\subsubsection{Explicit Synchronization}
Nvidia introduced cooperative group APIs~\cite{guide2022cuda} to provide a hierarchical of synchronizations in addition to thread block synchronization from P100 (2016). Among them, the new grid level synchronization provides additional choice for program. Zhang et. al.~\cite{zhang2020study} shows that latencies of these APIs are acceptable that would allow practical use.

\subsubsection{Asynchronous Shared Memory Copy.}
A100 (2020) further introduced APIs~\cite{guide2022cuda} to copy data from global memory to shared memory, without blocking. Martin et al.~\cite{svedin2021benchmarking} demonstrated that this API benefits low-arithmetic intensity kernels. 


%% file: LISTING/samplestencils.tex
\lstset{
 	language = C++, breaklines = true, breakindent = 10pt, lineskip={-1pt}, basicstyle = \rmfamily\scriptsize, commentstyle = {\itshape \color[cmyk]{1,0.4,1,0}}, classoffset = 0, keywordstyle = {\bfseries \color[cmyk]{0,1,0,0}}, stringstyle = {\ttfamily \color[rgb]{0,0,1}}, frame = trbl, framesep=0pt, numbers = left, stepnumber = 1, xrightmargin=12pt, xleftmargin=0pt, numberstyle = \tiny, tabsize = 1, captionpos = t, directivestyle={\color{black}},  emph={int,char,double,float,unsigned, int3, float4, float2}, emphstyle={\color{blue}},
}
\lstset{escapeinside={<@}{@>}}

\begin{figure}[t]
\centering
\begin{minipage}[c]{0.5\textwidth}
\begin{lstlisting}[caption = {Pseudocode for 1D 3-Point Jacobian Stencil}, label = Fig:1d3p]
for(int i=0; i<N; i++)
    out[i]=a*in[i-1]+b*in[i]+c*in[i+1];
\end{lstlisting}
\end{minipage}
\begin{minipage}[c]{0.5\textwidth}
\begin{lstlisting}[caption = {Pseudocode for 2D 5-Point Jacobian Stencil}, label = Fig:2d5p]
for(int i=0; i<N; i++)
    for(int j=0; j<M; j++)
        out[i][j]=a*in[i-1][j]+b*in[i][j]+c*in[i+1][j]
                +d*in[i][j-1]]+e*in[i][j+1];
\end{lstlisting}
\end{minipage}
\begin{minipage}[c]{0.5\textwidth}
\begin{lstlisting}[caption = {Pseudocode for 3D 7-Point Jacobian Stencil}, label = Fig:3d7p]
for(int i=0; i<N; i++)
    for(int j=0; j<M; j++)
        for(int k=0; k<L; k++){
            out[i][j][k]= a*in[i-1][j]  [k];
            out[i][j][k]+=b*in[i]  [j]  [k];
            out[i][j][k]+=c*in[i+1][j]  [k];
            out[i][j][k]+=d*in[i]  [j-1][k];
            out[i][j][k]+=e*in[i]  [j+1][k];
            out[i][j][k]+=f*in[i]  [j]  [k-1]; 
            out[i][j][k]+=g*in[i]  [j]  [k+1];}
\end{lstlisting}
\end{minipage}
\end{figure}

%% file: 03_EBISU.tex
\pagebreak
\section{EBISU: High Performance Temporal Blocking at Low Occupancy}\label{sec:ebisu}
\revision{In this section we give an overview of our temporal blocking method: EBISU (Figure~\ref{fig:overview} gives an overview)} The design of EBISU follows two main principles: minimal parallelism that would saturate the device \revision{(the Minimal Parallelism step in Figure~\ref{fig:overview})}, and scalability in using resources \revision{(the Implementation step in Figure~\ref{fig:overview})}. \revision{Additionally, EBISU relies on a comprehensive analysis for implementation decisions (the pink steps in Figure~\ref{fig:overview}).}

\subsection{Saturating the Device at Minimal Parallelism}
In EBISU we first tune the parallelism exposed in the kernel to find the minimal combination of occupancy and instruction level parallelism that would saturate the device. The minimal occupancy that we aim for in this paper is $12.5\%$ since further reducing the occupancy for memory-bound codes can start to regress the performance~\cite{nvidiaa100}. 
We aim to minimize resources used for increasing the locality. We use Little's Law to identify the minimum parallelism (occupancy) in the code (discussed in Section~\ref{sec:parallelismneed}). We point out that readers can also rely on auto-tuning tools to empirically figure out the minimal parallelism~\cite{kerneltuner,shende2006tau,rawat2019optimizing}.


\subsection{Scaling the Use of Resources}
Despite the relatively large amount on-chip resources, there is a lack in design principles that are able to scale up to take advantage of the large on-chip resources in temporal blocking. We thereby build on a set of existing optimizations to drive a resource-scalable scheme for increasing locality \revision{(Section~\ref{sec:implementations})}.

\revision{\subsection{Implementation Decisions}
We base the decision for implementing EBISU on our analysis for the practical attainable performance (Section~\ref{sec:model}). The main utility of this analysis is to decide whether to implement a device tile (Section~\ref{sec:gridornot}), and the parameterization of spatial and temporal blocking (Section~\ref{sec:parameters})).
\subsection{Fine-Tuning}
After identifying the ideal tiling scheme and parameterization, implementation, we fine-tune the kernel to extract additional performance. For instance, we tune the temporal blocking depth (Section~\ref{sec:dsrdept}).}

%% file: 04_Implementation.tex
\revision{
\section{Efficiently Scaling the Use of Resources}\label{sec:implementations}
}
\input{04_02_TileByTile}

\input{04_01_MultiQueue}
\input{04_03_Optimiations}

%% file: 04_02_TileByTile.tex
\begin{figure}[t]
\centering
\includegraphics[width=0.9\linewidth]{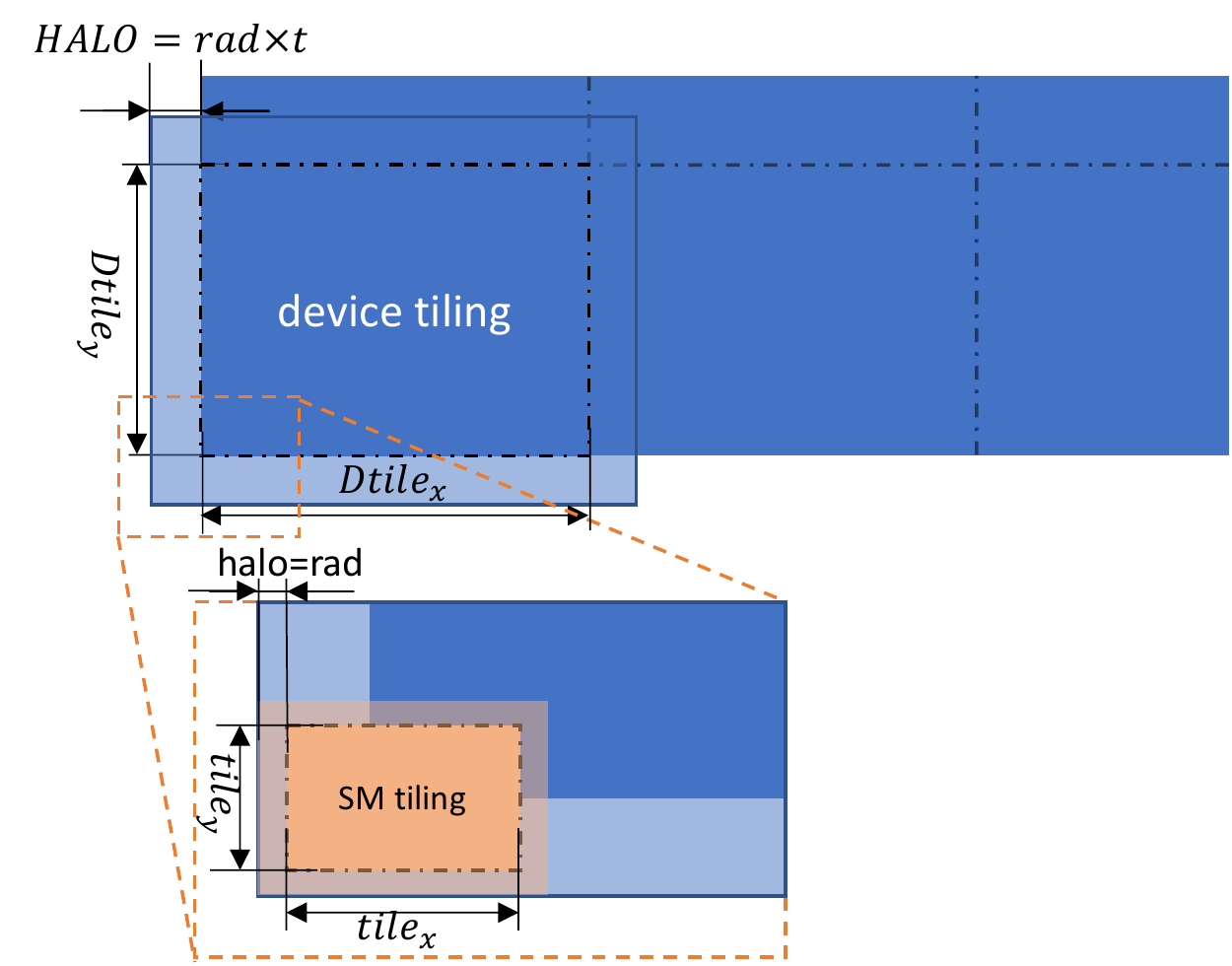}
\caption{\revision{2D Spatial tiling} at the GPU device level. }
\label{Fig:gtb}
\end{figure}

\input{LISTING/grid_level_temporal_blocking}

\revision{\subsection{One Tile At A Time}\label{sec:gtile}}
\revision{
Beyond the point where the GPU becomes saturated, the workload will inevitably be serialized. We intentionally introduce a method to serialize the execution of tiles, where each individual tile becomes large enough to saturate the GPU. We call this \emph{device tiling}. Alternatively, we can use tiles that are executed in parallel, yet each tile individually saturates a single streaming multiprocessor. We call this \emph{SM tiling}.
}


\revision{In device tiling,}
we tile the domain such that a single tile can scale up to use the entire \revision{on-chip} memory capacity of the GPU. Next, we let the tile reside in the \revision{on-chip} memory while updating the cells for \revision{a sufficient number of time steps} to amortize the initial loading and final storing overheads. We then store the final result for the tile on the device, and then we move to the next tile, i.e., the entire GPU is dedicated to computing only one single tile at any given time. Figure~\ref{Fig:gtb} shows how we do \revision{spatial tiling} at the device level. \revision{We assume $tile_x\times tile_y$ to be the thread block tile configuration and $Dtile_x\times Dtile_y$ to be the device tile configuration. Thus, $(tile_x+halo\cdot2)\times(tile_y+halo\cdot2)$ is the total on-chip memory consumed at the stream multiprocessor level. 
$(Dtile_x+HALO\cdot2)\times(Dtile_y+HALO\cdot2)$ is the total on-chip memory consumed at the device level, where $HALO=rad\cdot t$}. \revision{Additionally,} figure~\ref{fig:spblk}.c shows the dependency between thread blocks that we need to resolve. We use the bulk synchronous parallel (BSP) model to exchange the halo region and CUDA's grid level barrier for synchronization. We transpose the halo region that originally did not coalesce to reduce the memory transactions. \revision{Note that device tiling is an additional layer on top of SM tiling. Figure~\ref{Fig:gtb} shows an example of 2D spatial tiling at device level, and Listing~\ref{listing:gtb} presents the pseudocode of a 2D 5-point Jacobian stencil with device level spatial tiling.}

%% file: LISTING/grid_level_temporal_blocking.tex
\lstset{
 	language = C++, breaklines = true, breakindent = 10pt, lineskip={-1pt}, basicstyle = \rmfamily\scriptsize, commentstyle = {\itshape \color[cmyk]{1,0.4,1,0}}, classoffset = 0, keywordstyle = {\bfseries \color[cmyk]{0,1,0,0}}, stringstyle = {\ttfamily \color[rgb]{0,0,1}}, frame = trbl, framesep=0pt, numbers = left, stepnumber = 1, xrightmargin=12pt, xleftmargin=0pt, numberstyle = \tiny, tabsize = 1, captionpos = t, directivestyle={\color{black}},  emph={int,char,double,float,unsigned, int3, float4, float2}, emphstyle={\color{blue}},
}
\lstset{escapeinside={<@}{@>}}

\begin{figure}[t]
\centering
\begin{minipage}[c]{0.5\textwidth}
\begin{lstlisting}[caption = {Pseudocode for 2D 5-Point Jacobian stencil device level spatial tiling.}, label = listing:gtb]
void __global__ void device_2d5pt(...){...
    //data is loaded from on-chip memory ocm_in 
    //store data to ocm_out
    //ocm range tile_x, and tile_y;
    for(int s=0; s<t; s++){
        for(int l_y=0; l_y<tile_y; l_y+=1){
            for(int l_x=0; l_x<tile_x; l_x+=blockDim.x){
                ocm_out[i][j]=a*ocm_in[i-1][j]+b*ocm_in[i][j]
                             +c*ocm_in[i+1][j]+d*ocm_in[i][j-1]
                             +e*ocm_in[i][j+1];}
        __syncthreads();
        push_halo_to_neighbor(ocm_out[][],global_memory);
        grid.sync();
        swap(ocm_out,ocm_in);
        pull_halo_from_neighbor(ocm_in[][],global_memory);
        __syncthreads();
    }
...}
\end{lstlisting}
\end{minipage}
\end{figure}

%% file: 04_01_MultiQueue.tex
\begin{figure}[ht]

\begin{mdframed}[linecolor=white]
\centering
\includegraphics[width=\linewidth]{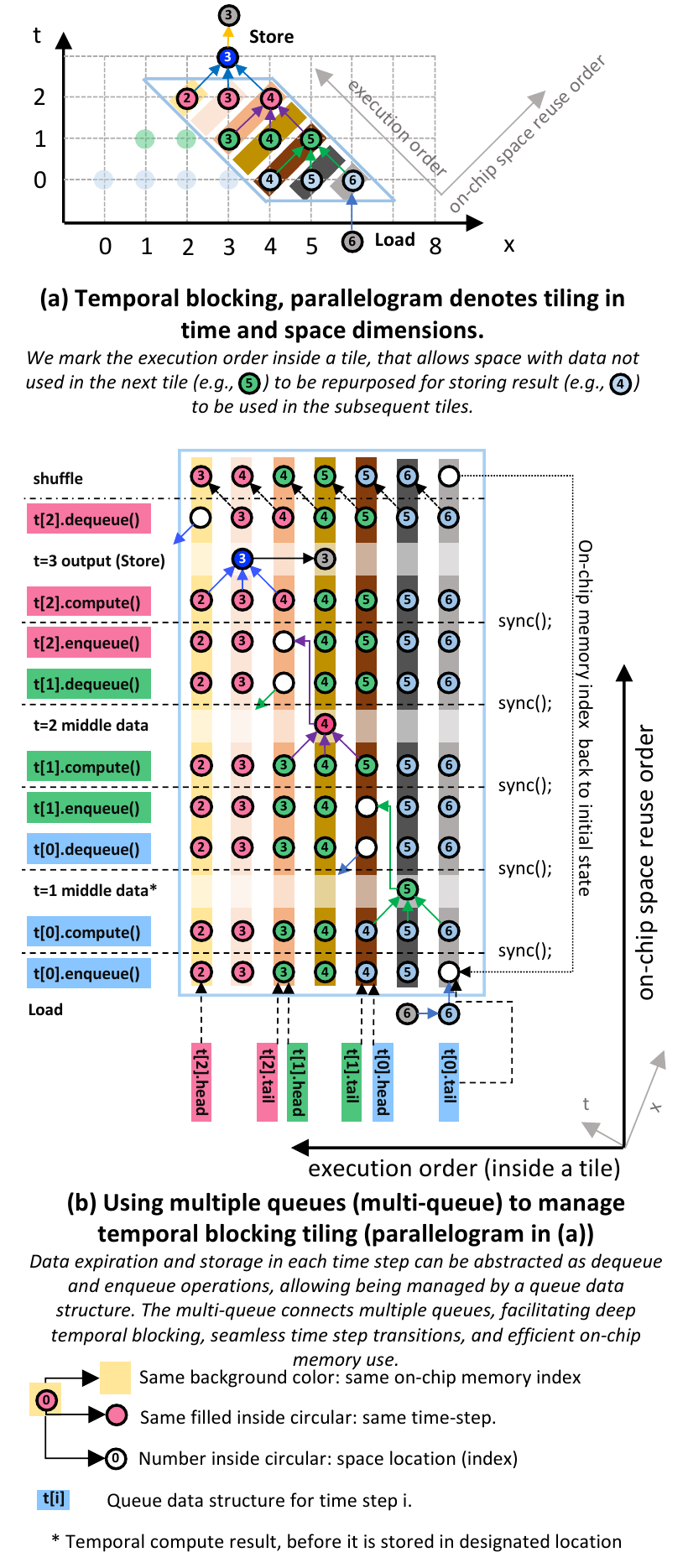}
\end{mdframed}
\caption{The multi-queue data structure enables efficient temporal blocking tiling: a 1D 3-Point Jacobian stencil with a depth of 3 as an example. Figure (a) illustrates streaming with a parallelogram that we process in Figure (b). Figure (b) illustrate how queue data structure can enhance the processing of the tiling depicted in Figure (a). The execution order and data reuse are marked in both figures.}
\label{Fig:mqs}
\end{figure}

\input{LISTING/1d3ptmultiqueue}

\input{LISTING/1d3ptcircularmultiqueue}
\revision{\subsection{Circular Multi-Queue}\label{sec:cycle}}
\revision{EBISU aims to scale up resource usage. One way to achieve this goal is to scale up to very deep temporal blocking.
In this section, we introduce a simple data structure that enables the efficient management of very deep temporal blocking: namely, \emph{circular multi-queue}. We elaborate on our design by first introducing \emph{multi-queue} for streaming (Section~\ref{sec:mq}), and then we describe the implementation of the \emph{circular multi-queue} (Section~\ref{sec:cmq}). 
\subsubsection{Multi-Queue}\label{sec:mq}
 We use the 1D 3-Point Jacobian stencil (Listing~\ref{Fig:1d3p}) to illustrate our implementation. Streaming is a typical method to implement temporal blocking. Figure~\ref{Fig:mqs}.a demonstrates an example of streaming. The parallelogram in the figure represents the tiling in time and spatial dimensions that we process in Figure~\ref{Fig:mqs}.b. The process of each time step can be abstracted as two functions: \textit{enqueue} and \textit{dequeue}, which are standard methods in a queue data structure. We additionally add \textit{compute} for stencil computation. As such, we manage each time step with a queue data structure. Next, we link queues in different time steps together, to become a {multi-queue} data structure. The data structure description and the pseudocode for multi-queue is in Listing~\ref{LIST:mqs}. 
 
{{Multi-queue} facilitates seamless transitions between time steps. The dequeue operation (data expiration) for the current time step runs concurrently with the enqueue operation for the next time step. After the execution of a single tile, we reset the multi-queue to its initial state - a process we refer to as 'shuffle'. A standard method of conducting a shuffle involves shifting values to their designated locations, as demonstrated in lines 24-27 of Listing~\ref{LIST:mqs}.}

It is important to note that although we base our analysis on a 1D stencil example in this section, it can be simply extended to 2D or 3D stencils by replacing the 1D circular points (domain cells) in Figure~\ref{Fig:mqs} to 1D lines (corresponding to 2D stencils) or 2D planes (corresponding to 3D stencils), or even the device tiles discussed in Section~\ref{sec:gtile}. In the device tiling situation, the sync(); function should be replaced by device (grid) level synchronization. Additionally, we can trade the concurrently processed domain cells for additional instruction level parallelism (ILP), which might be required by the parallelism setting (discussed in Section~\ref{sec:parallelismneed}).
}
\subsubsection{Circular Multi-Queue}\label{sec:cmq}
We further adapt the multi-queue to be circular. We wrap the tail of time step 0 and the head of the deepest time step together.
\revision{
We detail the implementation of different circular multi-queue we use as follows:
}


\noindent\textbf{Shifting Addresses:}
In this scheme, we only copy the index to the same place {\revision after processing a tile (at the 'shuffle step')}. 

\noindent\textbf{Computing Address:}
Shifting addresses is the simplest way to manage the circular data structure. However, shifting can create a long chain of dependencies as the address \revision{range} increases. An alternative solution is to compute the target address \revision{(Listing~\ref{LIST:cmqs} line 7-8.)}. The modulo operation is one of the solutions; however, this operator is time consuming. Instead, we extend the ring index to be $range=2^n, n\in \mathbb{Z}^+$. In this case, we have $index\%range=index\&(range-1)$. This consumes additional space \revision{(Listing~\ref{LIST:cmqs} line 22)}. 



%% file: LISTING/1d3ptmultiqueue.tex
\lstset{
 	language = C++, breaklines = true, breakindent = 10pt, lineskip={-1pt}, basicstyle = \rmfamily\scriptsize, commentstyle = {\itshape \color[cmyk]{1,0.4,1,0}}, classoffset = 0, keywordstyle = {\bfseries \color[cmyk]{0,1,0,0}}, stringstyle = {\ttfamily \color[rgb]{0,0,1}}, frame = trbl, framesep=0pt, numbers = left, stepnumber = 1, xrightmargin=12pt, xleftmargin=0pt, numberstyle = \tiny, tabsize = 1, captionpos = t, directivestyle={\color{black}},  emph={REAL,index,int,char,double,float,unsigned, int3, float4, float2}, emphstyle={\color{blue}},
}
\lstset{escapeinside={<@}{@>}}

\begin{figure}[t]
\centering
\begin{minipage}[c]{0.5\textwidth}
\begin{lstlisting}[caption = {Pseudocode for applying naive multi-queue data structure to a 1D 3-Point Jacobian stencil with temporal blocking depth of 3.}, label = LIST:mqs]
struct Queue { 
    REAL* d; //data array
    index tl; //tail
    index hd; //head
    Queue(REAL*data,index head,index tail):
        d(data),hd(head),tl(tail){}
    REAL dequeue(){}//Automatically accomplished by shuffle
    void enqueue(REAL input){d[tl]=input;}
    REAL compute()
        {return a*d[hd]+b*d[hd+1]+c*d[hd+2];}//1d3pt stencil 
}; 
template<int depth>
struct MultiQueue{//Multi-queue data structure
    REAL* d; //data array
    index hds[depth]; //head of queues
    index r;//range of multi-queue
    index q_r;//range of single queue, reserved for lazy streaming
    MultiQueue(REAL*data, index range, index queue_range):
        d(data),r(range),q_r(queue_range)
        {for(t=0; t<depth; t++)hds[t]=queue_range-q_r*s;}
    MultiQueue(REAL*data, index range):
        MultiQueue(data,range,2){}
    Queue operator[](int t) {return Queue(d,hds[t],hds[t]+q_r);}
    void shuffle(){//default, move data
        sync();
        for(int i=0; i<r-1; i++)d[i]=d[i+1];
        sync();
    }
}
#define RANGE (7)
__global__ void 1d3ptstencil(REAL* input, REAL* output,...){...
    REAL buffer[RANGE];
    MultiQueue t<3>(buffer, RANGE, 2);
    for(...){...
        t[0].enqueue(Load(input));
        sync();
        for(s=0; s<3-1; s++){
            tmp=t[s].compute();
            sync();
            t[s+1].enqueue(tmp);
            //Do t[s].dequeue() t[s+1].enqueue() 
            sync();
        }
        Store(ouput[],t[3-1].compute()...);
        ...
        t.shuffle();//shuffle head and tail index for next tiling
    ...}
}
\end{lstlisting}
\end{minipage}
\end{figure}

%% file: LISTING/1d3ptcircularmultiqueue.tex
\lstset{
 	language = C++, breaklines = true, breakindent = 10pt, lineskip={-1pt}, basicstyle = \rmfamily\scriptsize, commentstyle = {\itshape \color[cmyk]{1,0.4,1,0}}, classoffset = 0, keywordstyle = {\bfseries \color[cmyk]{0,1,0,0}}, stringstyle = {\ttfamily \color[rgb]{0,0,1}}, frame = trbl, framesep=0pt, numbers = left, stepnumber = 1, xrightmargin=12pt, xleftmargin=0pt, numberstyle = \tiny, tabsize = 1, captionpos = t, directivestyle={\color{black}},  emph={REAL,index,int,char,double,float,unsigned, int3, float4, float2}, emphstyle={\color{blue}},
}
\lstset{escapeinside={<@}{@>}}

\begin{figure}[t]
\centering
\begin{minipage}[c]{0.5\textwidth}
\begin{lstlisting}[caption = {Pseudocode for applying circular multi-queue data structure, which inhirits the structure described in Listing~\ref{LIST:mqs}.}, label = LIST:cmqs]
index mod(index a,index r){return (r&(r-1))==0)?a&(r-1):a%r;}
struct Circular_Queue: public Queue {
//Circular queue inhirit from queue
    index r;
    Circular_Queue(REAL*data,index head,index tail,index range,): 
        d(data),hd(head),r(range),tl(tail){}
    REAL compute()//override 1d3pt stencil 
        {return a*d[hd]+b*d[mod((hd+1),r)]+c*d[mod((hd+2),r)];}
}; 
template<int depth>//Circular multi-queue inhirit from multi-queue
struct Circular_MultiQueue: public MultiQueue<depth>{
    Circular_MultiQueue(REAL*data, index range)
        :Multi-queue<depth>(data,range){}
    Circular_MultiQueue(REAL*dat, index ran, index q_ran)
        :MultiQueue<depth>(dat,ran,q_ran){}
    Circular_Queue operator[](int t)
        {return Circular_Queue(d,hds[t],mod(hds[t]+2,r),r);}
    void shuffle(){//Override shuffle for computing address schema
        for(int i=0; i<r; i++)hds[i]=mod((hds[i]+1),r);
    }
}
#define RANGE (8)
...//kernel code unchange
\end{lstlisting}
\end{minipage}
\end{figure}

%% file: 04_03_Optimiations.tex
\begin{figure}[t]
\centering
\includegraphics[width=0.9\linewidth]{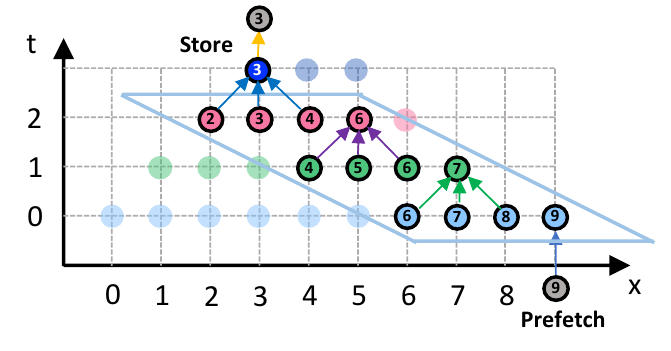}
\caption{{Lazy streaming} for temporal blocking. 1D 3-Point Jacobian stencil with depth=3 as an example. Notations are the same as Figure~\ref{Fig:mqs}.}
\label{Fig:lazystreaming}
\end{figure}
\input{LISTING/1d3ptlazystreaming}
\revision{\subsection{Optimizations}}
\subsubsection{Prefetching}\label{sec:prefetch}
Prefetching is a well-documented optimization. Readers can refer to~\cite{rawat2019optimizing} for hints. The new asynchronous shared memory copy API \revision{offers another approach} for prefetching, with a trade-off of requiring additional shared memory space for buffering.

\subsubsection{Lazy Streaming}\label{sec:lazystreaming}
\revision{The naive implementation showed in Figure~\ref{Fig:mqs} and Listing~\ref{LIST:mqs} clearly suffers from the overhead of frequent synchronization.}
We propose \emph{lazy streaming} \revision{to alleviate this type of overhead}. The basic idea is that we delay the processing of a domain cell until all domain cells required to update the current cell are already updated. Otherwise, we would pack the planes that include the current domain cell and cache it in on-chip memory. As Figure~\ref{Fig:lazystreaming} shows, the computation of \textit{location 3} is postponed until the three points of the previous time steps have been updated. 
 
The benefit of using {lazy streaming} is not significant in 1D stencils. In 2D or 3D stencils, we replace the points in Figure~\ref{Fig:lazystreaming} with 1D or 2D planes for 2D or 3D stencils. The planes usually involve inter-thread dependency, which makes synchronizations unavoidable (warp shuffle when using registers for locality~\cite{chen2019versatile, sat} or thread block synchronization when using shared memory for locality~\cite{maruyama2014optimizing}). \revision{when applying \emph{device tiling} (Section~\ref{sec:gtile}), device (grid) level synchronization becomes unavoidable, and it has higher overhead in comparison to thread block synchronizations.} 
\revision{As illustrated in Listing~\ref{LIST:lazy},} 
{lazy streaming} can ideally reduce the synchronization to one synchronization per \revision{tile}. The benefit of lazy streaming comes from the number of synchronization it reduced. 

\revision{It's worth noting that double-buffering~\cite{DBLP:conf/cgo/MatsumuraZWEM20,rawat2018domain} can be viewed as a special case of {lazy streaming} when only a single queue evolved. }


\subsubsection{Redundant Register Streaming}\label{sec:redundant}

\revision{
The above discussions, which do not specify the on-chip memory type, can apply to both shared memory-based and register based implementations. However, there is one exception: the circular multi-queue cannot be implemented with register arrays since register addresses cannot be determined at compile time.
}

At low occupancy, we obtain a large number of registers and shared memory per thread. Therefore, by reducing the occupancy, we can afford to redundantly store intermediate data in both the registers and the shared memory. Streaming w/ caching in shared memory is discussed in STENCILGEN~\cite{rawat2018domain}. Streaming w/ caching in the registers is discussed in AN5D~\cite{DBLP:conf/cgo/MatsumuraZWEM20}. We benefit from both components by caching in both shared memory and registers. We can reduce shared memory access times to their minimum by using registers first (in comparison to AN5D) and reducing the necessary synchronizations when using only shared memory (in comparison to STENCILGEN). \revision{Additionally, due to data being mostly redundant, we can tune to reduce resource usage in either part of registers or shared memory to reduce the resource burden.}

%% file: LISTING/1d3ptlazystreaming.tex
\lstset{
 	language = C++, breaklines = true, breakindent = 10pt, lineskip={-1pt}, basicstyle = \rmfamily\scriptsize, commentstyle = {\itshape \color[cmyk]{1,0.4,1,0}}, classoffset = 0, keywordstyle = {\bfseries \color[cmyk]{0,1,0,0}}, stringstyle = {\ttfamily \color[rgb]{0,0,1}}, frame = trbl, framesep=0pt, numbers = left, stepnumber = 1, xrightmargin=12pt, xleftmargin=0pt, numberstyle = \tiny, tabsize = 1, captionpos = t, directivestyle={\color{black}},  emph={REAL,index,int,char,double,float,unsigned, int3, float4, float2}, emphstyle={\color{blue}},
}
\lstset{escapeinside={<@}{@>}}

\begin{figure}[t]
\centering
\begin{minipage}[c]{0.5\textwidth}
\begin{lstlisting}[caption = {\revision Pseudocode for applying lazy streaming to a 1D\\ 3-Point Jacobian stencil with temporal blocking depth of 3.}, label = LIST:lazy]
//Lazy streaming kernel code 
__global__ void 1d3ptstencil_lz(REAL* input, REAL* output,...){...
    REAL buffer[16];//more space for buffering
    Circular_MultiQueue t<3>(buffer, 16, 3);
    for(...){...
        t[0].enqueue(Load(input));//prefetch
        for(s=0; s<3-1; s++){
            tmp=t[s].compute();
            t[s+1].enqueue(tmp);
        }
        Store(ouput[],t[3-1].compute()...);
        ...
        t.shuffle();//shuffle head and tail index for next tiling
        sync();//One sync per tile
        ...
    }
}
\end{lstlisting}
\end{minipage}
\end{figure}

%% file: 05_AttainablePerformance.tex
\revision{
\section{Practical Attainable Performance}\label{sec:model}
In this section, we analyze the practical attainable performance of temporal blocking by incorporating an the overhead analysis (we derive valid proportion $\mathbb{V}$ from overhead analysis in Section~\ref{sec:overhead}) to a roofline-like model~\cite{ofenbeck2014applying,kim2011performance} that predicts the attainable performance ($\mathbb{P}$ in Section~\ref{sec:bottlenecks}).}
We project the practical attainable performance $\mathbb{PP}$ as:
\begin{equation}\footnotesize
    \mathbb{PP}= \mathbb{P}\times{\mathbb{V}}
\end{equation}

\revision{The model proposed in this section serves as a guide for implementation design choices in Section~\ref{sec:designchoice}.}

\subsection{{Attainable Performance}} 
\label{sec:bottlenecks}
We use the giga-cells updated per second (GCells/s) as the metric for stencil performance~\cite{DBLP:conf/cgo/MatsumuraZWEM20,chen2019versatile}. We consider three pressure points in a stencil kernel: double precision ALUs, cache bandwidth (i.e., shared memory bandwidth in this paper), and device memory bandwidth (GPU global memory in this paper). Note that registers could also be a pressure point in extreme cases of very high order stencils (outside the scope of this paper).

Assuming that the global memory bandwidth is $\mathbb{B}_{gm}$, the shared memory bandwidth is $\mathbb{B}_{sm}$, and the compute speed is $\mathbb{THR}_{cmp}$, the total access time is $\mathbb{A}_{gm}$ and $\mathbb{A}_{sm}$ for global memory and shared memory, respectively. The total amount of computation is $\mathbb{A}_{cmp}$. The memory access time per cell is ${a}_{gm}$ and ${a}_{sm}$ for global memory and shared memory, respectively; flops per cell is ${a}_{cmp}$. The total number of cells in the domain of interest is $\mathbb{D}_{gm}$, $\mathbb{D}_{sm}$ and $\mathbb{D}_{gm}$ for global memory, shared memory, and computation, respectively. The size of the cell (in Bytes) per cell is $\mathbb{S}_{Cell}$. We can compute the runtime of using each component to be:
\begin{equation}\footnotesize
\begin{split}
    \mathbb{T}_{gm}&= \frac{\mathbb{A}_{gm}}{\mathbb{B}_{gm}}\times \mathbb{S}_{Cell}=\frac{{a}_{gm}\times\mathbb{D}_{gm}}{\mathbb{B}_{gm}}\times \mathbb{S}_{Cell}
\end{split}
\end{equation}
\begin{equation}\footnotesize
\begin{split}
    \mathbb{T}_{sm}&= \frac{\mathbb{A}_{sm}\times t}{\mathbb{B}_{sm}}\times\mathbb{S}_{Cell}= \frac{{a}_{sm}\times\mathbb{D}_{sm}\times t}{\mathbb{B}_{sm}}\times \mathbb{S}_{Cell}
\end{split}
\end{equation}
\begin{equation}\footnotesize
\begin{split}
    \mathbb{T}_{com}&= \frac{\mathbb{A}_{cmp}\times t}{\mathbb{THR}_{cmp}}= \frac{{a}_{cmp}\times\mathbb{D}_{cmp}\times t}{\mathbb{THR}_{cmp}}\\
\end{split}
\end{equation}

The total runtime of the stencil is projected as:
\begin{equation}\footnotesize
    \mathbb{T}_{stencil}= \max(\mathbb{T}_{gm},\mathbb{T}_{sm},\mathbb{T}_{cmp})
\end{equation}

The component $c$ is the bottleneck if it satisfies: 
\begin{equation}\footnotesize
    \mathbb{T}_{c}= \mathbb{T}_{stencil}
\end{equation}

\revision{
We project the attainable performance $\mathbb{P}$ as:
\begin{equation}\footnotesize
    \mathbb{P}= \frac{\mathbb{D}_{all}\times{t}}{\mathbb{T}_{stencil}}
\end{equation}
}

Normally, we consider $\mathbb{D}_{all}=\mathbb{D}_{sm}=\mathbb{D}_{gm}=\mathbb{D}_{cmp}$. However, this is a case-by-case factor that depends on the implementation, i.e., when applying \emph{device tiling} $\mathbb{D}_{gm}\neq\mathbb{D}_{sm}$. 

\subsection{Overheads}~\label{sec:overhead}
In this section, we discuss the overheads of different spatial blocking methods used in this paper: 

\subsubsection{SM Tiling}\label{sec:overheadOverlappedTilling}
The main overhead of SM tiling is related to redundant computation in halo. Only a portion of the computation is valid. This valid portion is related to both the spatial and temporal block sizes and the radius of the stencil. In 2D stencils, we have: 
\begin{equation}\footnotesize
    \mathbb{V}= \frac{tile_x-t\times rad}{tile_x}
\end{equation}
In 3D stencils, we have:
\begin{equation}\footnotesize
    \mathbb{V}_{SMtile}= \frac{(tile_x-t\times rad)\times(tile_y-t\times rad)}{tile_x\times tile_y}
\end{equation}
Accordingly, we have:
\begin{equation}\footnotesize
    \mathbb{PP}_{SMtile}= \mathbb{V}_{SMtile}\times\mathbb{P}
\end{equation}
\subsubsection{Device Tiling}\label{sec:gridtileoverhead}
The main overhead of the device level tiling is related to the overhead of synchronization. Only a portion of the \revision{runtime} is valid. The valid portion depends on the runtime of the stencil ($\mathbb{T}_{stencil}$), the time required for device level synchronization ($\mathbb{T}_{Dsync}$) \revision{and the number of synchronization times per tile $n$ (applying \emph{lazy streaming} (Section~\ref{sec:lazystreaming}) reduces $n$ to 1):}
\begin{equation}\footnotesize
    \mathbb{V}_{Dtile}= \frac{\mathbb{T}_{stencil}}{\mathbb{T}_{stencil}+\mathbb{T}_{Dsync}\times n}
\end{equation}
Accordingly, we have:
\begin{equation}\footnotesize
    \mathbb{PP}_{Dtile}= \mathbb{V}_{Dtile}\times\mathbb{P}
\end{equation}
To quantify the overhead, we followed the research of Zhang et al.~\cite{zhang2020study} to test the overheads. The device (grid) level synchronization overhead in A100 is : $\mathbb{T}_{Dsync}=1.2us$. 

%% file: 06_ImplementationDecisions.tex
\input{TABLE/designDecision}
\section{EBISU: Analysis of Design Choices}\label{sec:designchoice}
In this section, \revision{we provide a comprehensive analysis to justify our design choices.}
The analysis is targeted at the A100 GPU, while it can be generalized to any GPU platform by adjusting the model parameters (Table~\ref{tab:designDecision} summarizes our findings on design choices). 

\revision{We use 2D 5-Point (Listing~\ref{Fig:2d5p}) to represent 2D stencils, and 3D 7-Point (Listing~\ref{Fig:3d7p}) to represent 3D stencils for the discussions in this section. Table~\ref{tab:domain} shows the detailed parameters of both stencils.}

\input{06_01_minimalpar}

\input{06_04_DesiredDepth}
\input{06_02_ToGSyncOrNotTo}
\input{06_03_DeeperWider}

%% file: TABLE/designDecision.tex
\begin{table*}[t]
    \centering
    \caption{\revision{Design choices for EBISU.}}
    {%
    \footnotesize
    \begin{tabular}{|c|c|c|c|c|c|}\hline
\textbf{Type}                     &\textbf{Parallelism Combination } &\textbf{ SM Tiling}  &\textbf{Device Tiling }  &\textbf{Temporal Blocking Strategy} &\textbf{Circular Multi-Queue}\\
                &\textbf{ ($N_{threads}\times ILP$)} & \textbf{ ($tile_{x}\times tile_{y}$)} & & &\\\hline 
\textbf{2D stencils}              & $256\times4$&  $256\times4$&       --    & Deep enough to shift the bottleneck &Compute              \\\hline
\textbf{3D stencils }  & $256\times4$ &$32\times32$&     $(12\times6)$    & As deep as possible&Shifting             \\\hline
    \end{tabular}
    }
    \label{tab:designDecision}
\end{table*}

%% file: 06_01_minimalpar.tex
\subsection{Minimum Necessary Parallelism}\label{sec:parallelismneed}
The analysis of this section is an extension of Volkov's work on low occupancy at high performance~\cite{volkov2010better}. We also generalize the analysis by building on Little's law. Little's saw uses latency $\mathbb{L}$ and throughput $\mathbb{THR}$ to infer the concurrency $\mathbb{C}$ of the given hardware:
\begin{equation}\footnotesize
     \mathbb{C}=\mathbb{L}\times\mathbb{THR}
\end{equation}

The latency $\mathbb{L}$ of an instruction can be gathered by common microbenchmarks~\cite{wong2010demystifying,mei2016dissecting}. The throughput $\mathbb{THR}$ of instructions is available in Nvidia's CUDA programming guide~\cite{guide2022cuda} and documents~\cite{nvidiaa100}.

As long as the parallelism $\mathbb{PAR}$ provided by the code is larger than the concurrency provided by the hardware, we consider that the code saturates the hardware:

\begin{equation}\footnotesize
     \mathbb{PAR}\ge\mathbb{C}
\end{equation}

There are two ways of providing parallelism: number of threads ($N_{threads}$) and Instruction Level Parallelism ($ILP$). So, we have:

\begin{equation}\footnotesize
     \mathbb{PAR}={N_{threads}}\times{ILP}
\end{equation}

Unlike Volkov's analysis, instead of maximizing the parallelism with the combination of $ILP$ and $N_{threads}$, we aim to find a minimal combination of $N_{threads}$ and $ILP$ that saturates the device:

\begin{equation}\footnotesize
     {N_{threads}}\times{ILP}=\mathbb{PAR}\ge\mathbb{C}=\mathbb{L}\times\mathbb{THR}
\end{equation}

To maintain a certain level of parallelism, we can reduce the occupancy ($N_{threads}$) and increase $ILP$ simultaneously. We reduce the occupancy to the point that it will not increase the resources per thread block. In the current generation of GPUs (A100), reducing the occupancy of memory-bound kernels to less than $12.5\%$ will not increase the available register per thread~\cite{guide2022cuda}. So, we set our aim conservatively at $Occupancy=12.5\%$ or $N_{threads}=256$. 

In this research, we focus on double precision global memory access, shared memory access, and DFMA, all of which are the basic operations in stencil computation. Based on our experimentation, $ILP=4$ and $Occupancy=12.5\%$ ($N_{threads}=256$) provide enough parallelism for all three operations. We set this as a basic parallelism combination for our implementation. Note that the numbers above may vary for other GPUs, yet the analysis still holds.

%% file: 06_04_DesiredDepth.tex
\revision{
\subsection{Desired Depth}\label{sec:dsrdept}
We use the attainable performance analysis (Section~\ref{sec:bottlenecks}) to infer the desired depth.}
We aim at determining a sufficiently \revision{deep} temporal blocking size to shift the bottleneck. 

In this study, we are less concerned with whether the bottleneck shifts to computation or cache bandwidth. To simplify the discussion, we assume that the optimization goal is shifting the bottleneck from global memory to shared memory. This assumption is true for most of the star-shaped stencils~\cite{DBLP:conf/cgo/MatsumuraZWEM20}. Accordingly, we have: 
\begin{equation}\footnotesize
    \frac{{a}_{sm}\times t}{\mathbb{B}_{sm}}\times{\mathbb{D}_{sm}} \geq \frac{{a}_{gm}}{\mathbb{B}_{gm}}\times\mathbb{D}_{gm}
    \label{eqt:basidassume}
\end{equation}

\subsubsection{\revision{Case Study: 2D 5-Point Jacobian Stencil (representing stencils w/o device tiling)}}\label{sec:2d5ptdesire}
Ideally, we have $\mathbb{D}_{sm}=\mathbb{D}_{gm}$. In A100, $\mathbb{B}_{gm}=1555$ GB/s, $\mathbb{B}_{sm}=19.49$ TB/s. In our 2D 5-point implementation, $\mathbf{a}_{gm}=2$ (assuming perfect caching), $\mathbf{a}_{sm}=4$. According to Equation~\ref{eqt:basidassume}, we have $t\geq6.3$. In $t=7$, we measured the performance of $440$ GCells/s. We can \revision{fine-tune to} achieve slightly better performance at $t=12$, where we measured $482$ GCells/s. There is only a $10\%$ difference in performance. The slight inaccuracy might come from the fact that, on average, the global memory accesses per data point is not perfectly cached. 
\revision{
\subsubsection{\revision{Case Study: 3D 7-Point Jacobian Stencil (representing stencils w/ device tiling)}}\label{sec:3d7ptdesire}
In device tiling 3D 7-point stencil, $\mathbb{D}_{gm}$ must also include the halo region between thread blocks. As such, we have:
\begin{equation}\footnotesize
    \mathbb{D}_{gm}=(tile_x\times tile_y)+(tile_x+ tile_y)\times2\times t \times rad
\end{equation}

We intend to determine a $t$ that satisfies: 
\begin{equation}\footnotesize
    \frac{{a}_{sm}\times\mathbb{D}_{sm}\times t}{\mathbb{B}_{sm}}>
    \frac{{a}_{gm}\times\mathbb{D}_{gm}}{\mathbb{B}_{gm}}
\end{equation}

We assume that $tile_x=tile_y=32$. We have $a_{sm}=4.5$, $a_{gm}=2$. So we can get $t>18.34$. In this situation, the on-chip memory per thread block desired for EBISU is $352$ KB, which exceeds the capacity of A100 ($164$ KB).

}

%% file: 06_02_ToGSyncOrNotTo.tex
\revision{\subsection{Device Tiling or SM Tiling?}\label{sec:gridornot}
Device tiling trades redundant computation for device level synchronization.}
In this section, we focus on: in EBISU, the performance implications of using one single tile per device (w/ device level synchronization). \revision{By comparing the practical attainable performance with the version that is not using one single tile per \revision{device} (w/o device level synchronization).  
}
\subsubsection{Case Study: 2D 5-Point Jacobian Stencil}\label{sec:cost2d5pt}
In 2d5pt, we have $\mathbb{T}_{stencil}=\mathbb{T}_{sm}$ for the overlapped tilling and the device level tiling. We simplify the discussion by defining a valid proportion $\mathbb{V}$, i.e., the updated output after excluding the halo. The higher the valid proportion, the higher the performance $\mathbb{P}$. 
In overlapped tiling, for 2d5pt we have $t=7$ \revision{(Section~\ref{sec:2d5ptdesire})} and $rad=1$. So $\mathbb{V}_{SMtile}\approx95\%$

For device level tiling, we can go as deep as $t=15$. So, we have: $\mathbb{T}_{sm}=2.05us$. Because $\mathbb{T}_{Dsync}=1.2us$. Accordingly, we have $\mathbb{V}_{Dtile}={\mathbb{T}_{sm}}/{(\mathbb{T}_{sm}+\mathbb{T}_{Dsync})}\approx63\%$. 

So, we have: $\mathbb{V}_{Dtile}\ll\mathbb{V}_{SMtile}$.

For 2D stencils of other shapes, we get: 
\begin{equation}\footnotesize
    \mathbb{PP}_{Dtile}(2D)\ll\mathbb{PP}_{SMtile}(2D)
\end{equation}

\revision{As a result, in 2D stencils, the overhead of thread block level overlapped tiling is negligible, making device tiling less beneficial.} This result stands true for all 2D stencils we studied in A100. 

\subsubsection{Case Study: 3D 7-Point Jacobian Stencils} \label{sec:cost3d7pt}
In 3d7pt, we cannot shift the bottleneck to shared memory in overlapped (within acceptable overhead) or device tiling. We need to compare the Practical Attainable Performance in both cases to judge.%

We have $\mathbb{V}_{SMtile}={(34-2\times rad\times t)^2}/{34^2}$. In 3d7pt, we have $rad=1$, $t=3$, $\mathbb{V}_{SMtile}\approx77\%$. In $t=3$, we have $\mathbb{P}_{SMtile}=292$ GCells/s, and $\mathbb{PP}_{SMtile}\approx225$ GCells/s.

On the other hand, for device tiling, we can go as deep as $t=8$, so we have $\mathbb{L}(gm)=2.42$ . Because $\mathbb{T}_{Dsync}=1.2$ us. So, $ \mathbb{V}_{Dtile}\approx67\%$ GCells/s. In $t=8$ we have $\mathbb{P}_{Dtile}=365$ GCells/s. Accordingly, we have $\mathbb{PP}_{Dtile}\approx244$ GCells/s.

So, we have: $\mathbb{PP}_{Dtile}>\mathbb{PP}_{SMtile}$ on a 3d7pt stencil.

We measured, for instance, $151$ GCells/s for \revision{w/o} device tiling and $197$ GCells/s for \revision{w/} device tiling. The experiment results is consistent with the analysis (for 3D stencils of other shapes as well): 
\begin{equation}\footnotesize
    \mathbb{PP}_{Dtile}(3D) > \mathbb{PP}_{SMtile}(3D)
\end{equation}

\revision{As a result, for 3D stencils, the overhead of thread block level overlapped tiling is so significant that it prohibits the temporal blocking implementation from going  deeper.} 
This result stands true for all 3D stencils we studied in A100. 

\revision{Based on the analyses above, in EBISU, we only implement device tiling for 3D stencils. The analysis in the following section is built on top of this decision. }



%% file: 06_03_DeeperWider.tex
\revision{
\subsection{Deeper or Wider?}\label{sec:parameters}
As the capacity of on-chip memory is limited, there is a trade-off between increasing the width of spatial blocking and increasing the depth of temporal blocking. In this section, we discuss our heuristic for we use for parameter selection in EBISU. 
}

\subsubsection{\revision{Case Study: 2D 5-Point Jacobian Stencil}}\label{sec:2d5ptana}
Firstly, as Section~\ref{sec:cost2d5pt} showed, \revision{the overhead of 2D 5-Point Jacobian Stencil is negligible. Additionally, according to Section~\ref{sec:2d5ptdesire}, in theory, at depth $t=7$, we shift the bottleneck from global memory to shared memory. 
}

\revision{As such, after the bottleneck is shifted, we aim at wider spatial blocking to reduce the overhead of overlapped tilling as is discussed in Section~\ref{sec:overheadOverlappedTilling}. Yet, we still need to consider the implementation simplicity. For example, we choose a tiling of size $tile_x=256$ instead of $tile_x=328$, since the latter is hard to implement in CUDA. 
}.

\subsubsection{\revision{Case Study: 3D 7-Point Jacobian Stencil}}\label{sec:3d7ptana}

For simplicity, we assume that the very first plane loaded and the last plane stored have already been amortized. Then, for global memory access, we only focus on the halo region. According to Equation~\ref{eqt:basidassume}, we have:
\begin{equation}\footnotesize
    \frac{tile_x\times tile_y \times{a}_{sm}}{\mathbb{B}_{sm}} > \frac{(tile_x+tile_y)\times2\times{a}_{gm}\times rad}{\mathbb{B}_{gm}}
\end{equation}

We assume that $tile_y=tile_x$. So, we can get: 
\begin{equation}\footnotesize
    tile_y=tile_x >{\frac{4\times {a}_{gm}\times \mathbb{B}_{sm}}{{a}_{sm}\times\mathbb{B}_{gm}}}\times rad
\end{equation}

In our 3d7pt implementation, ${a}_{gm}=2$, ${a}_{sm}=4.5$. We have $tile_y=tile_x\geq 22.3$. For \revision{implementation} convenience, we use $32\times32$ (also fitted to the Minimal Necessary Parallelism that saturates the device as Section~\ref{sec:parallelismneed} discussed). \revision{As such, after} the spatial tiling is large enough for overlapping halo region, we then run the temporal blocking as deep as possible to amortize the overhead of using device (grid) level synchronization. 

%% file: 07_evaluation.tex
\section{Evaluation}
\input{TABLE/domaininfo}

\input{TABLE/temporalblockingsize}
\begin{figure*}[t]
\centering
\includegraphics[width=\textwidth]{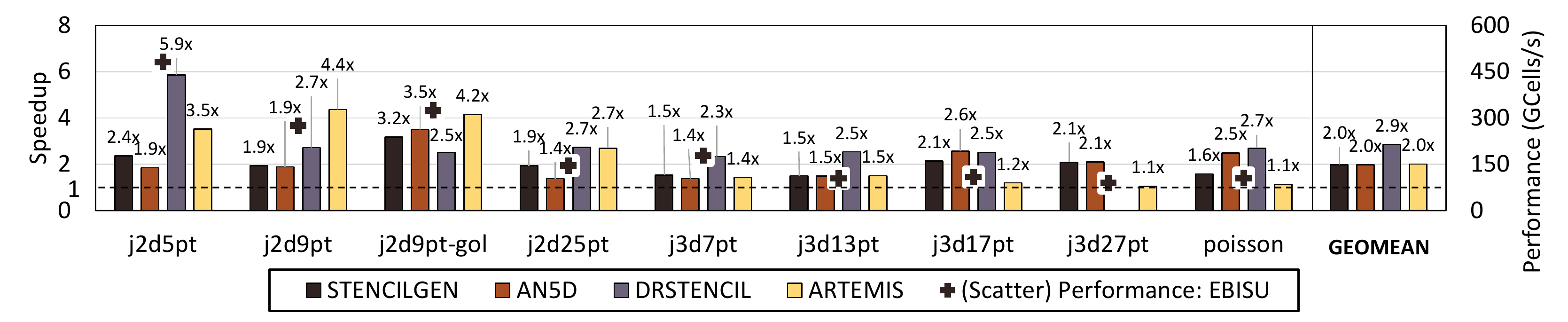}
\caption{\label{fig:evaluation}Speedup of EBISU over the state-of-the-art temporal blocking implementations. We also plot the performance of EBISU (right Y-axis plotted as '$+$' ticks). 
}
\end{figure*}

\begin{figure}[t]
\centering
\includegraphics[width=\linewidth]{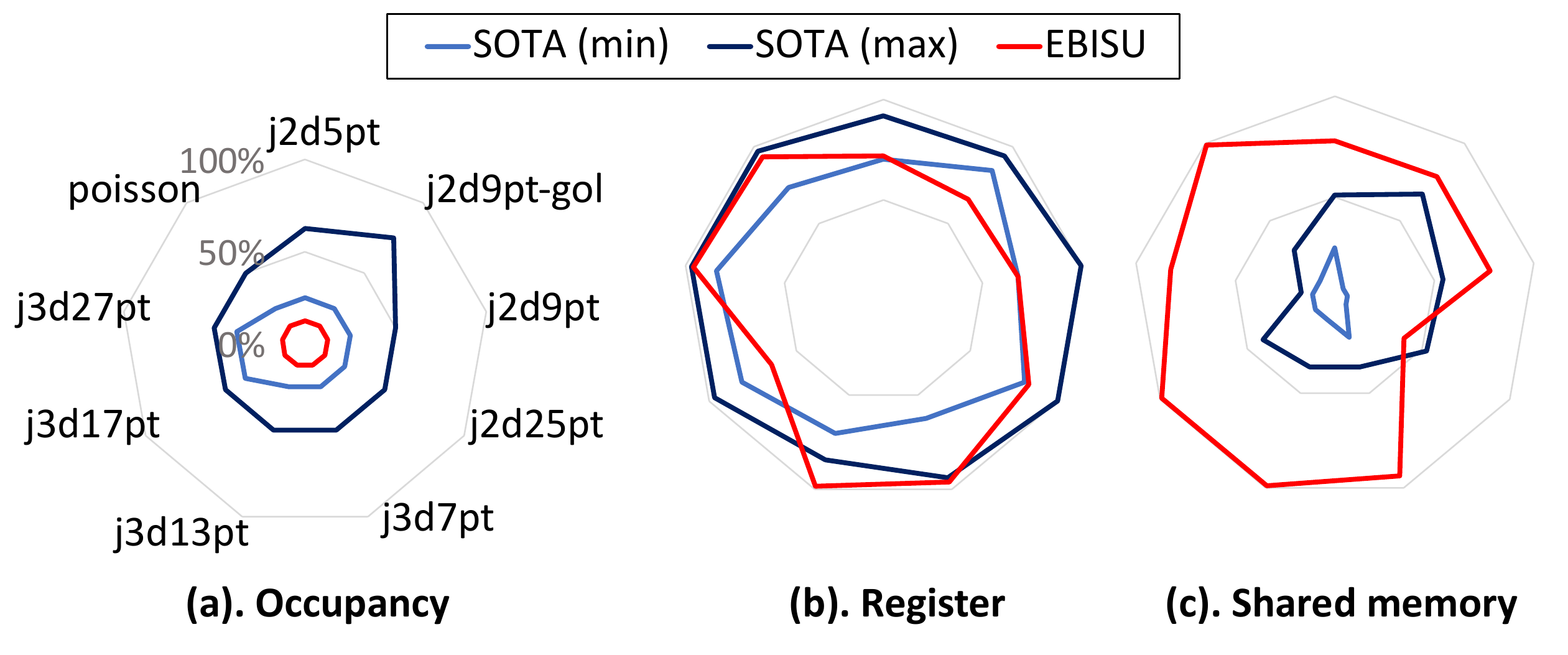}
\caption{\label{fig:heatmap} \revision{Percent of} occupancy achieved and resources used (registers and shared memory) \revision{for EBISU and SOTA libraries among all stencil benchmarks}.
}
\end{figure}

We experiment on a wide range of 2D and 3D stencils (listed in Table~\ref{tab:domain}). The test data are generated by STENCILGEN~\cite{rawat2018domain}. 
\revision{We evaluate the benchmarks on an NVIDIA A100-PCIe GPU device(host CPU: Intel Xeon E5-2650).
\subsection{Compile Settings of EBISU}
The code is compiled with NVCC-11.5 (CUDA driver V11.5.119) and gcc-4.8.5, using flags {-rdc=true -Xptxas "-v " -std=c++14}. We only generate code for A100 architecture~\footnote{setting CUDA\_ARCHITECTURES "80" in CMAKE}. 
"--rdc=true" flag is necessary for enabling grid level synchronization, so we set it by default. We use c++14 features, so we add "-std=c++14" flag. -Xptxas "-v" is set to gather information on registers. 
}
\subsection{Evaluation Setup}
\subsubsection{Domain Size}

We used the domain sizes listed in Table~\ref{tab:domain} for EBISU, comparable to those used in the literature~\cite{chen2019versatile,bondhugula2017diamond,Rawat:2018:ROS:3200691.3178500}.

\subsubsection{Warm-Up and Timing}
For all experiments, we do warm-up iterations and then use GPU event APIs to measure one kernel run. We repeat this process ten times and report the \revision{peak}.

\subsubsection{Depth of Temporal Blocking} 
We only evaluate a single kernel. Therefore, the total number of time steps is equal to the depth of temporal blocking of each implementation in each stencil benchmark. We summarize the depth of temporal blocking in Table~\ref{tab:tblk}. 


\input{07_01_SOTAs}

\input{07_02_BreakDown}

%% file: TABLE/domaininfo.tex
\begin{table}[t]
    \centering
    \caption{Stencil benchmarks. Readers can refers to ~\cite{rawat2018domain,DBLP:conf/cgo/MatsumuraZWEM20} for details description. \revision{We also include ideal shared memory access times per cell, $a_{sm}$, when applying redundant register streaming (w/ RST) and without it (w/o RST) in the table.}}
    {%
    \footnotesize
    \begin{tabular}{|l|c|c|c|c|}\hline
\textbf{{\footnotesize{Stencil}}}  &\textbf{Domain Size} &$a_{sm}$& $a_{sm}$\\  
\textbf{{\footnotesize{[Order, FLOPs/Cell]}}}  & & \textbf{w/o RST}&\textbf{ w/ RST}\\  \hline 
\textbf{j2d5pt} [1,10]   & $8352^2$          &6&4            \\
\textbf{j2d9pt} [2,18]      & $8064^2$       &10&6         \\
\textbf{j2d9pt-gol} [1,18]       & $8784^2$  &10&4                \\
\textbf{j2d25pt (gaussian)} [2,25]     & $8640^2$  &26&6             \\\hhline{|=|=|=|=|}
\textbf{j3d7pt (heat)} [1,14]& $2560\times288\times384$ &8&4.5   \\
\textbf{j3d13pt} [2,26]  & $2560\times288\times384$    &14&7 \\
\textbf{j3d17pt} [1,34]  & $2560\times288\times384$    &18&5.5\\
\textbf{j3d27pt} [1,54]  & $2560\times288\times384$    &28&5.5\\
\textbf{poisson} [1,38] & $2560\times288\times384$     &20&5.5\\\hline
    \end{tabular}
    }
    \label{tab:domain}
\end{table}

%% file: TABLE/temporalblockingsize.tex
\begin{table}[t]
    \centering
    \caption{Depth of temporal blocking for each stencil implementations in this evaluation.}
    {%
   \footnotesize
    \begin{tabular}{|l|c|c|c|c|c|}\hline
\textbf{Type} & \scriptsize{\textbf{STENCILGEN}}  &\ {\textbf{AN5D} }&  \scriptsize{\textbf{DRSTENCIL}}&\scriptsize {\textbf{ARTEMIS}}& {\textbf{EBISU}}\\  \hline 
\textbf{j2d5pt      }& 4  & 10  & 3  &12&  12           \\
\textbf{j2d9pt }    & 4   & 5   & 2  &6&     8     \\
\scriptsize{\textbf{j2d9pt-gol}} &4    &7    & 2  &6&  6            \\
\textbf{j2d25pt}   & 2   &  5  & 2  &3&   4      \\\hhline{|=|=|=|=|=|=|}
\textbf{j3d7pt }    &4    &  6  &3  &3&8 \\
\textbf{j3d13pt}    & 2   &  4  &2  &1&5\\
\textbf{j3d17pt}    & 2   & 3   &2  &2&6\\
\textbf{j3d27pt}    & 2   & 3   & - &2&5          \\
\textbf{poisson}    & 4   &3    & 2  &  2&6\\\hline
    \end{tabular}
    }
    \label{tab:tblk}
\end{table}


%% file: 07_01_SOTAs.tex
\subsection{{Comparing with State-Of-The-Art Implementations}}
We compare EBISU with the state-of-the-art temporal blocking implementations AN5D~\cite{DBLP:conf/cgo/MatsumuraZWEM20} and STENCILGEN~\cite{rawat2018domain}, and the state-of-the-art auto-tuning tools ARTEMIS~\cite{rawat2019optimizing} and DRSTENCIL~\cite{you2021drstencil}.

\subsubsection{\revision{Setting up State-Of-The-Art Libraries}}
We use the domain sizes reported by each library in the original paper (not adversely change domain sizes). We assume that the libraries can achieve reasonably good performance in the setting used in the original paper. For example, in 2D stencils, AN5D used $16384^2$, while STENCILGEN used $8192^2$. ARTEMIS did not report 2D stencils; we used the same setting as STENCILGEN. Details can be obtained from the original papers~\cite{DBLP:conf/cgo/MatsumuraZWEM20,rawat2018domain,rawat2019optimizing,you2021drstencil}. 

\revision{As for timing and warm-up.} AN5D's original code already does the warm-up, so we use the default setting. We use the same host warm-up and timer function as EBISU to test the kernel performance for STENCILGEN, ARTEMIS, and DRStencil. 

\revision{The detailed settings are listed as follows:}

\noindent\textbf{STENCILGEN}
We used the codes for AD/AE appendix~\cite{stencilgenadae}
of the original paper. We do not change anything inside the kernel. 

\noindent\textbf{AN5D}
AN5D is a code auto-generator tool. We only used the code already generated in their code~\cite{an5dadae}
. We port the makefile system to A100 and iterate over all generated codes to find the one with the highest performance for each stencil benchmark. The original code did not include some stencil benchmarks we use. We use the implementations with similar memory access patterns to represent them: gaussian (box2d2r), j3d7pt (star3d1r), j3d13pt (star3d2r), j3d17pt (j3d27pt) and poisson (j3d27pt).

\noindent\textbf{DRSTENCIL}
DRSTENCIL~\cite{you2021drstencil} is also an auto-tuning tool. We use the benchmark in the codebase~\cite{drstencilcode}
. In the paper, the authors included only the implementations of the j3d7pt stencil in the range of 3D stencils. We extend their j3d7pt stencil setting to other 3D stencils for comparison. However, with the j3d7pt setting, DRSTENCIL was unable to generate runnable code in j3d27pt. We report the kernel with the peak performance among the policies that DRSTENCIL iterated over.

\noindent\textbf{ARTEMIS}
ARTEMIS is an auto-tuning tool. We use the benchmark in the codebase~\cite{artemiscode}
. We replaced the profiler nvprof (deprecated) with ncu. ARTEMIS~\cite{rawat2019optimizing} only provides samples for 3d7pt and 3d27pt. We extend 3d7pt to all star-shape stencils (including heat and 2d star-shape stencils) and 3d27pt to all box-shape stencils (including poisson, 3d17pt and 2d box-shape stencils). We report the kernel with the peak performance among the policies that ARTEMIS iterated over.

\subsubsection{Performance Comparison}
Figure~\ref{fig:evaluation} shows the speedup of EBISU over state-of-the-art temporal blocking implementations. EBISU shows a clear performance advantage over all of the state-of-the-art temporal blocking libraries, i.e., STENCILGEN and AN5D. It is also faster than the state-of-the-art auto-tuning tool DRSTENCIL and ARTEMIS. EBISU achieves a geomean speedup of over $2.0$x when comparing with each state-of-the-art. When comparing EBISU with the best state-of-the-art in each stencil, EBISU achieves a geomean speedup of $1.49$x.

\subsubsection{Resources}

We additionally report occupancy and the resources used for all the benchmarks with the ncu profiler (Figure~\ref{fig:heatmap}). EBISU is able to use the on-chip resources efficiently despite its low occupancy ($12.5\%$). It is worth noting that, as Table~\ref{tab:tblk} shows, EBISU usually has deeper temporal blocking. However, EBISU does not show significantly higher register pressure than other implementations. EBISU can, on average, do temporal blocking $1.3$x deeper than the deepest state-of-the-art implementations. But only use $87\%$ of the registers compared to the most register-consuming state-of-the-art equivalent kernel.   


%% file: 07_02_BreakDown.tex
\begin{figure*}[t]
\centering
\includegraphics[width=\linewidth]{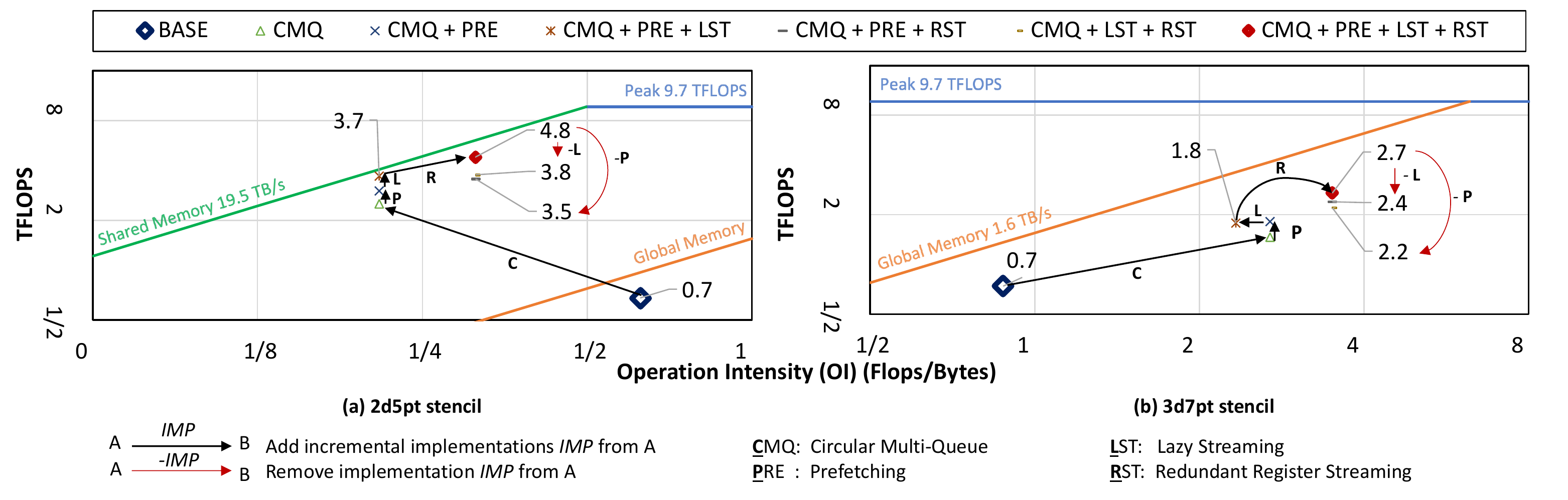}
\caption{\revision Roofline plots for different implementations in Section~\ref{sec:implementations}. We plot 2D 5-Point Jacobian stencil implementations to represent 2D stencils and 3D 7-Point Jacobian stencil implementations to represent 3D stencils. The black arrows link the incremental implementations from a $BASE$ implementations. The 3D $BASE$ applies device tiling (Section~\ref{sec:gtile}). The $LST$ refers to thread block level {lazy streaming}. {Device tiling} without {lazy streaming} will be extremely slow as can be inferred in Section~\ref{sec:gridtileoverhead}.}
\label{Fig:breaksroofline}
\end{figure*}

\subsection{Performance Breakdown}
The remarkable speedup achieved by EBISU in comparison to other SOTA methods can be attributed to a fundamental shift in GPU programming principles. While existing SOTAs typically focus on constraining resources to enhance parallelism, EBISU constrains parallelism to optimize resource utilization. This novel approach enables the implementation of resource-scalable schemes, which ultimately contribute to EBISU's performance.

In this section, we provide a detailed explanation of how the optimizations proposed in earlier sections impact the performance of EBISU. To demystify their effects, we present case studies involving 2D 5-Point Jacobian stencils (representing 2D stencils) and 3D 7-Point Jacobian stencils (representing 3D stencils). Figure~\ref{Fig:breaksroofline} displays the roofline plot of various implementations, with the black arrow indicating the incremental implementation of each scheme.

For the roofline analysis, we report the performance as measured in TFLOPS (teraflops). Table~\ref{tab:domain} shows the relationship between TFLOPS and GCells/s metrics.

\subsubsection{BASE} 
{
The BASE implementation refers to the approach that applies minimal parallelism analysis, as discussed in Section~\ref{sec:parallelismneed}. In this phase, we prepare the necessary resources for EBISU. It is important to note that in the case of the 3D 7-Point stencil, the BASE implementation already incorporates \emph{device tiling}, similar to the approach employed in the existing research of PERKS~\cite{zhang2023persistent}.
}

\subsubsection{Circular Multi-Queue (CMQ)} 
CMQ is a foundation for deep temporal blocking. As Figure~\ref{Fig:breaksroofline} shows, in 2D stencils, we increase the depth of temporal blocking to move the bottleneck from global memory to shared memory. In 3D stencils, due to the shared memory's limited capacity, we only move the Operation Intensity (OI) from left to right. Either way, we move the OI such that we increase the attainable performance shown in the roofline model.

\subsubsection{Prefetching (PRE)} 

As Figure~\ref{Fig:breaksroofline} shows, the PRE scheme has the effect of moving the roofline plot towards the attainable bound. However, it does not directly impact the attainable bound itself.

\subsubsection{Lazy Streaming (LST)}
The LST scheme aims to reduce synchronizations by using long buffers. By default, we employ LST to minimize device level synchronizations. This section specifically focuses on the impact of LST on reducing thread block synchronizations.
As illustrated in Figure~\ref{Fig:breaksroofline}.a, applying LST to the 2D 5-point stencil brings its performance closer to the attainable bound. However, in the case of the 3D stencil, as shown in Figure~\ref{Fig:breaksroofline}.b, applying LST may harm performance. This is primarily due to the global memory still being the bottleneck, and the additional on-chip memory space required by LST implementation leads to a shallower temporal blocking. This results in a leftward shift in the OI, which consequently reduces the attainable performance.
It is worth nothing that in the final version of EBISU, disabling LST for the 3D 7-point stencil allows for a doubling of the temporal blocking depth, from $t=8$ to $t=16$, leading to a performance increase from 2.7 TB/s to 2.9 TB/s. However, when excluding the redundant halo, the performance dips from 2.4 TB/s to 2.3 TB/s. Therefore, this result has been excluded from the discussion

\subsubsection{Redundant Register Streaming (RST)} 
RST's primary goal is to cut down shared memory access time (refer to Table~\ref{tab:domain}). By doing so, we can shift the roofline plot closer to the compute bound from left to right when shared memory is the bottleneck (as shown in Figure~\ref{Fig:breaksroofline}.a). Also, we leverage RST to cache a portion of the tiling, which helps reduce the amount of data cached in shared memory. This enables us to achieve deeper temporal blocking and move the roofline plots closer to the compute bound from left to right, when global memory remains the bottleneck (as shown in Figure~\ref{Fig:breaksroofline}.b).
\subsubsection{Relations Between Optimizations}
The PRE and LST optimizations have the effect of improving performance and bringing it closer to the attainable bound. The RST optimization is designed to shift the roofline plots to the right, to increase the attainable bound. Red arrows in Figure~\ref{Fig:breaksroofline} clearly shows that disabling either of these optimizations results in a degradation of performance.

\subsubsection{Practical Attainable Performance}
In 2D 5-point stencil, we achieved $4.8$ TFLOPS ($80\%$ of the attainable bound). In 3D 7-point stencil, we achieved $2.7$ TFLOPS  ($50\%$ of the attainable bound). The big gap is due to the omission of the overheads in roofline model. As we consider overhead in our model (Section~\ref{sec:model}), we achieved $88\%$ and $80\%$ of $\mathbb{PP}$ in 2D 5-point and 3D 7-point stencils respectively. A model that considers the overheads can model the practical attainable performance better. As such, this model contributing to the decision-making also benefits the performance of EBISU. 

%% file: 08_relatedwork.tex
\section{Related works}

Apart from the tiling optimizations we covered in Section~\ref{sec:stencilback}, there are many stencil optimizations that are architecture-specific. For example, vectorization~\cite{zhao2019exploiting,yuan2021temporal,henretty2011data}; cache optimizations on CPUs~\cite{wellein2009efficient,tang2015cache,malas2015multicore,akbudak2020asynchronous}. For GPUs~\cite{rawat2016effective,grosser2014hybrid,Holewinski:2012:HCG:2304576.2304619}, Chen et al. proposed an execution model on top of the shuffle operation on GPU~\cite{chen2019versatile}; Liu et al. uses tensor cores to accelerate low precision stencils~\cite{liu2022toward}. Rawat et al. also summarized optimizations that can be used in stencil optimization, i.e., streaming, unrolling, prefetching~\cite{rawat2019optimizing}, and register reorder~\cite{Rawat:2018:ROS:3200691.3178500}.

State-of-the-art implementations are usually built on top of multiple optimizations. For example, wavefront diamond blocking~\cite{malas2015multicore} is built on top of vectorization, cache optimization, streaming, and diamond tiling, STENCILGEN~\cite{rawat2018domain} is built on top of shared memory optimization, streaming, and N.5D tiling. 

But combining different optimizations is tedious for implementation. Many researches focus on autocode generation using on domain specific language~\cite{rawat2018domain,maruyama2011physis,zhao2018delivering}, or compiler-based approaches~\cite{verdoolaege2013polyhedral,ragan2013halide}. Some optimizations, especially those related to registers, are difficult to implement manually. Matsumura et al. implemented AN5D~\cite{DBLP:conf/cgo/MatsumuraZWEM20} that generates codes using registers effectively. 

%% file: 09_conclusion.tex
\section{Conclusion and Future Work}
In this paper we propose, EBISU, a novel temporal blocking approach. EBISU relies on low occupancy and mapping on large tiles over the device. The freed resources are then used to improve the data locality. We compared EBISU with two state-of-the-art temporal blocking implementations and two state-of-the-art autotuning tools. EBISU constantly shows its performance advantage. It achieves a geomean speedup of $1.49$x over any of the top alternative state-of-the-art implementations for each stencil benchmark.

\revision{
This paper focuses on studying how modern GPU characteristics influence the optimization of temporal blocking stencils. Nevertheless, as EBISU proved effective, its optimization approach can be absorbed into production libraries like Halide~\cite{ragan2013halide} so that the end user can get the performance with minimal effort.
}

%% file: 10_ack.tex
\begin{acks}
This work was supported by JSPS KAKENHI under Grant Numbers JP22H03600 and JP21K17750. This work was supported by JST, PRESTO Grant Number JPMJPR20MA, Japan. This paper is based on results obtained from JPNP20006 project, commissioned by the New Energy and Industrial Technology Development Organization (NEDO).
This manuscript has been co-authored by UT-Battelle, LLC, under contract DE-AC05-00OR22725 with the US Department of Energy (DOE). The publisher acknowledges the US government license to provide public access under the DOE Public Access Plan (\url{http://energy.gov/downloads/doe-public-access-plan}/).
The authors wish to express their sincere appreciation to Jens Domke, Aleksandr Drozd, Emil Vatai and other RIKEN R-CCS colleagues for their invaluable advice and guidance throughout the course of this research. Finally, the first author would also like to express his gratitude to RIKEN R-CCS for offering the opportunity to undertake this research in an intern program. 
\end{acks}